\documentclass{article}
\usepackage[utf8]{inputenc}
\usepackage{amsmath}

\title{Landauer's princple for Fermionic field in one dimensional bag}
\author{Yu-Song Cao\footnote{caoyusong15@mails.ucas.ac.cn}$~^{1}$, YanXia Liu\footnote{yxliu-china@ynu.edu.cn}$~^{1}$, Rong Zhang\footnote{zhangrong@csrc.ac.cn}$~^{2}$}
\date{}

\begin{document}

\maketitle

\noindent $~^{1}$\small{School of Physics and Astronomy, Yunnan University, Kunming 650091, PR China}

\noindent $~^{2}$\small{Beijing Computational Science Research Center, Beijing 100193, People's Republic of China}

\begin{abstract}
    We study the Landauer's principle of an Unruh-DeWitt detector linearly coupled to Dirac field in $1+1$ dimensional cavity. When the initial state of the field is vacuum, we obtain the heat transfer and von Neumann entropy change perturbatively. For the thermal state, the heat transfer and entropy change are approximately obtained in the case where the interaction time is long enough and the Unruh-DeWitt detector is in resonance with one of the field mode. Compared to the real scalar field, we find the results of vacuum initial state differs solely from the helicity of the Dirac field and the distinguishablity of fermion and anti-fermion comes into play when the initial state is thermal. We also point out that the results for massless fermionic field can be obtained by taking the particle mass $m\rightarrow 0$. We find that in both cases satisfy Landauer's principle. 
\end{abstract}

\section{Introduction}

    One of the pillars the current knowledge about universe built upon is the quantum field theory, which gives an almost perfect description about the dynamics and interaction of the microscopic particles and is the foundation of many branches of the modern physic such as particle physics, condensed matter physics and quantum optics. Conventionally, the study of quantum field theory is focused on the dynamics and interaction. The pursue of the foundation of quantum physics requires us examining the quantum information concepts like measurement theory and entanglement on the quantum field theory level \cite{hu,measurement1,D96,vacuum,mutual,measure}. One of the most important application of this direction is the exploration of the nature of black hole entropy, where the corresponding microstates are believed to be in quantum nature, calls for a deeper and throughout discussion about the relation between gravity, quantum field theory, information and thermodynamcis.
    
    The research on the information aspects of quantum field theory are diverse and has already led to many astonishing results such as the violation of Bell's inequality of vacuum state \cite{werner1,werner2,A71,PL} and entanglement harvesting \cite{fermion,dce,harvest,harvest2,graviton}. To study the connection between information and thermodynamics there is Landauer's principle, which states that at least $k_{B}T\ln{2}$ energy is required to erase one bit of information \cite{landauer61,landauer96}, which provides a new perspective of studying the nature of black hole microstates \cite{entropy,daf2007}. The discussion on the implication of Landauer's principle on quantum field theory is a newly emerged topic. In \cite{wyj1,wyj2}, the authors discussed the Landauer's principle of an Unruh-DeWitt detector coupled to real scalar field through monopole momentum. In quantum theory, the field can be classified into two kinds: fermionic ones and bosonics ones, and they behaves very differently from each other. Thus, it is meaningful to study the Landauer's princple in fermionic field. 
    
    In this paper, we study the Unruh-DeWitt detector traveling through Dirac field in a $1+1$ dimensional cavity, where the cavity walls are modeled by MIT bag boundary condition \cite{soviet,1204}. The Dirac field is linearly coupled to Unruh-DeWitt detector whose strength is small thus allowing the use of perturbation theory. We obtain the heat transfer and entropy change are obtained in two cases, one is the Dirac field is prepared initially in vacuum state while another is the thermal initial state with long interaction period and the detector is in resonance with one of the field model. By comparing with real scalar field \cite{wyj1}, we find that for vacuum initial state, the heat transfer and entropy grow are modified only due to the helicity of the fermion and anti-fermion. While for thermal initial state, the results are modified also from the distinguishability of the fermion and anti-fermion. We point out that the results for a massless fermionic field can be obtained as $m\rightarrow 0$ limit of our results. The Landauer's principle is found to hold in both cases. This paper is organized as follows. In Sec.\ref{sec:2}, we demonstrate the set up of our system and provide the formula to be used later. The heat transfer and entropy change when the Dirac field is initially prepared in vacuum state are given in Sec.\ref{sec:3}. In Sec.\ref{sec:4}, we derive the formula of the thermal Dirac field state first, then we obtain the heat transfer and entropy change when the detector is in resonance with one of the field modes. Conclusion and discussion are made in Sec.\ref{sec:5}.
    
    Throughout this paper the unit are chosen as $\hbar=c=k_{B}=1$.

\section{Model Setup}\label{sec:2}

    In our model, we consider a Dirac field under the boundary condition
    \begin{equation}\label{eq:diraceq}
        \begin{split}
            &(i\gamma_{\mu}\partial^{\mu}-m)\psi(t,x)=0,\\
            &(e^{i\theta\gamma_{5}}+in^{\mu}\gamma_{\mu})\psi(t,x)|_{\Sigma}=0,
        \end{split}
    \end{equation}
    where $\mu=0,1$ and $\Sigma$ is the boundary of the cavity, which in this case are the points $x=0$ and $x=L$ and $L$ is length of cavity. $n_{\mu}$ is the normal vector of the boundary. The spacetime metric is $g_{\mu\nu}=diag(1.-1)$ and the notation $\gamma_{0}=\sigma_{1},\gamma_{1}=\sigma_{3},\gamma_{5}=\sigma_{2}$ is used, where $\sigma_{\alpha},\alpha=0,1,2,3$ denotes the Pauli matrices. The Dirac field can be mode expanded as
    \begin{equation}
        \psi(t,x)=\sum_{n}b(k_{n})e^{-i\omega_{n}t}f_{n}(x)+d^{\dagger}(k_{n})e^{i\omega_{n}t}h_{n}(x),
    \end{equation}
    where $n$ is mode number, $b(k_{n})$ and $d^{\dagger}(k_{n})$ are the annihilation operator of fermion and the creation operator of anti-fermion in mode $n$, and $\{f_{n}(x),h_{n}(x)\}$ are the normalized basis of solution of Dirac equation (\ref{eq:diraceq}), which is given as \cite{D105}
    \begin{equation}
        \begin{split}
            &f_{n}[x(t)]=N_{n}\left(
                \begin{array}{cc}
                    &\frac{\omega_{n}}{k_{n}}\sin{(k_{n}x)} \\
                    & \cos{(k_{n}x)}+\frac{m}{k_{n}}\sin{(k_{n}x)}
            \end{array}
            \right),\\
            &h_{n}[x(t)]=N_{n}\left(
                \begin{array}{cc}
                    &-\frac{\omega_{n}}{k_{n}}\sin{(k_{n}x)} \\
                    & \cos{(k_{n}x)}+\frac{m}{k_{n}}\sin{(k_{n}x)}
            \end{array}
            \right),\\
            &N_{n}=\sqrt{2}k_{n}^{2}[k_{n}^{2}(m+2L\omega_{n}^{2})+m\omega_{n}^{2}\sin^{2}{(k_{n}L)}]^{-\frac{1}{2}},
        \end{split}
    \end{equation}
    where the mode frequency is $\omega_{n}=\sqrt{k_{n}^{2}+m^{2}}$ and $N_{n}$ is the normalization constant, with which the field modes can be obtained satisfying transcendental equation \cite{D105}
    \begin{equation}
        \frac{m}{k_{n}}\sin{(k_{n}L)}+\cos{(k_{n}L)}=0.
    \end{equation}
    Note that by taking $m\rightarrow 0$ we obtain the formula for massless fermionic field straightforwardly.
    
    The  total Hamiltonian of our system consists of three parts $H=H_{f}+H_{D}+H_{int}$, where $H_{f},H_{D}$ and $H_{int}$ denotes the Hamiltonian of the free Dirac field inside the cavity, the free detector and the interaction between the detector and the Dirac field, respectively. In interaction picture,  $H_{0},H_{D}$ and $H_{int}$ take form of 
    \begin{equation}\label{eq:hamiltonian}
        \begin{split}
            &H_{f}=\sum_{n}\omega_{n}[b^{\dagger}(k_{n})b(k_{n})+d^{\dagger}(k_{n})d(k_{n})],\\
            &H_{D}=\Omega\sigma_{z},\\
            &H_{int}=\lambda(e^{i\Omega t}\sigma_{+}\bar{\psi}[x(t)]\Lambda(t)+e^{-i\Omega t}\sigma_{-}\bar{\Lambda}(t)\psi[x(t)]),
        \end{split}
    \end{equation}
    where $\bar{\psi}(t,x)=\psi^{\dagger}(t,x)\gamma^{0}$ and the Pauli matrices $\sigma_{z},\sigma_{\pm}$ act on the Hilbert space of the detector. The smear function $\Lambda(t)=\eta\chi(t)$, where $\eta$ is a Grassman number, $\chi(t)=1$ when $0<t<T$ and vanishes otherwise. The worldline of the Unruh-DeWitt detector is $x(t)$.
    
    The evolution operator is given by
    \begin{equation}\label{eq:evolution}
        U(t)=\mathcal{T}\exp{(-i\int_{0}^{t}dt'H_{int}(t')}),
    \end{equation}
    where $\mathcal{T}$ is the time-ordering symbol. Expand (\ref{eq:evolution}) with respect to $\lambda$, we have $U(t)=U^{(0)}(t)+U^{(1)}(t)+U^{(2)}(t)+\mathcal{O}(\lambda^{3})$, where $U^{(i)}(t)$ denotes the $i$-th order of the expansion. The initial density matrix $\rho_{0}$ evolves according to
    \begin{equation}
        \rho(t)=U(t)\rho_{0}U^{\dagger}(t).
    \end{equation}
    Thus the final density matrix $\rho(T)$ can be expanded with respect to $\lambda$ as $\rho_{T}=\rho^{(0)}_{T}+\rho^{(1)}_{T}+\rho^{(2)}_{T}+\mathcal{O}(\lambda^{3})$, where
    \begin{equation}
        \begin{split}
            &\rho^{(0)}_{T}=\rho_{0},\\
            &\rho^{(1)}_{T}=U^{(1)}(T)\rho_{0}+\rho_{0}U^{(1)\dagger}(T),\\
            &\rho^{(2)}_{T}=U^{(2)}(T)\rho_{0}+U^{(1)}(T)\rho_{0}U^{(1)\dagger}(T)+\rho_{0}U^{(2)\dagger}(T).
        \end{split}
    \end{equation}

\section{Vacuum state}\label{sec:3}

    In this section we study the situation where the Dirac field is initially prepared in vacuum state $|0\rangle$, which is a thermal state corresponding to the tempreture $T_{R}=0$. Thus the initial state of the whole system is $\rho_{0}=[(1-p)|-\rangle\langle-|+p|+\rangle\langle+|]\otimes|0\rangle\langle 0|$, where $|\pm\rangle$ corresponds to the excited and ground state of the Unruh-DeWitt detector, and $p$ is a real number for the possibility of the detector being in excited state. The heat transferred to the reservoir and entropy change of the systems are given by \cite{1306}
    \begin{equation}
        \begin{split}
            &\Delta Q=tr[H_{f}(\rho_{T,f}-\rho_{0,f})],\\
            &\Delta S=S(\rho_{0,D})-S(\rho_{T,D}),
        \end{split}
    \end{equation}
    where $S$ is von Neumann entropy. From (\ref{eq:hamiltonian}) we can see that only the diagonal elements of the reduced density matrix of the Unruh-DeWitt detector and Dirac field contribute to $\Delta Q$ and $\Delta S$. Since $U^{(1)}(t)$ is made up of terms like $\sigma_{+}d,\sigma_{+}b^{\dagger},\sigma_{-}b,\sigma_{-}d^{\dagger}$, its action upon $\rho_{0}$ will only result in non-diagonal elements in the reduced density matrices of the detector and Dirac field, which means $\rho^{(1)}(t)$ does not contribute to the heat transfer and entropy grow.
    
    Now we consider the the second order contribution $\rho^{(2)}_{T}$, With (\ref{eq:hamiltonian}) we obtain the relevant contributions are
    \begin{equation}
        \begin{split}
            &U^{(1)}(T)\rho_{0}U^{(1)\dagger}(T)\rightarrow\lambda^{2}\sum_{n}\bigg((1-p)|W_{n}|^{2}|+\rangle\langle+|\otimes|n\rangle\langle n|+p|V_{n}|^{2}|-\rangle\langle-|\otimes|\bar{n}\rangle\langle \bar{n}|\bigg),\\
            &U^{(2)}(T)\rho_{0}=\rho_{0}U^{(2)\dagger}(T)=-\frac{\lambda^{2}}{2}\sum_{n}\bigg((1-p)|W_{n}|^{2}|-\rangle\langle-|+p|V_{n}|^{2}|+\rangle\langle+|\bigg)\otimes|0\rangle\langle 0|,
        \end{split}
    \end{equation}
    where the rightarrow means only the diagonal terms are kept and $|n\rangle,|\bar{n}\rangle$ denotes one electron and positron state of mode $n$, respectively. And the notations
    \begin{equation}\label{eq:WV}
        \begin{split}
            &W_{n}=\int_{0}^{T}dt e^{-i(\Omega+\omega_{n})t}\bar{\eta}f_{n}[x(t)],\\
            &V_{n}=\int_{0}^{T}dt e^{i(\Omega-\omega_{n})t}\bar{h}_{n}[x(t)]\eta
        \end{split}
    \end{equation}
    are used. Thus the density matrix of field are obtained
    \begin{equation}
        \rho_{T,f}=\bigg(1-\lambda^{2}\sum_{n}[(1-p)|W_{n}|^{2}+p|V_{n}|^{2}]\bigg)|0\rangle\langle 0|+\lambda^{2}(1-p)\sum_{n}|W_{n}|^{2}|n\rangle\langle n|+\lambda^{2}p\sum_{n}|V_{n}|^{2}|\bar{n}\rangle\langle \bar{n}|,
    \end{equation}
    as well as the reduced density matrix of detector
    \begin{equation}\label{eq:rhod}
        \rho_{T,D}=(1-p-\delta p)|-\rangle\langle-|+(p+\delta p)|+\rangle\langle+|,
    \end{equation}
    where 
    \begin{equation}
        \delta p=\lambda^{2}\sum_{n}[(1-p)|W_{n}|^{2}-p|V_{n}|^{2}].
    \end{equation}
    Then we obtain the heat transfer to the Dirac field and the entropy grow of the detector
    \begin{equation}\label{eq:deltaqs}
        \begin{split}
            &\Delta S=\lambda^{2}\ln{\frac{1-p}{p}}\sum_{n}[p|V_{n}|^{2}-(1-p)|W_{n}|^{2}],\\
            &\Delta Q=\lambda^{2}\sum_{n}[(1-p)|W_{n}|^{2}+p|V_{n}|^{2}]\omega_{n}.
        \end{split}
    \end{equation}
    From (\ref{eq:deltaqs}) we can see that in this case the difference between heat transfer and entropy grow of real scalar field and fermionic field solely comes from the helicity of the Dirac field by adding the factors $\bar{\eta}f_{n}[x(t)]$ and $\bar{h}_{n}[x(t)]\eta$ in the integrand (\ref{eq:WV}). And we obtain that the heat transfer is always non-negative, which is in agreement with intuition that being the ground state of $H_{f}$, the Dirac vacuum can only absorb energy while the entropy of the detector can either grow or decrease. In this case the Landauer's principle is always fulfilled due to $T_{R}=0$.
    
\section{Thermal state}\label{sec:4}
    In this section we consider the case where the Dirac field is initially in thermal state. The Hilbert space of the Dirac field can be factorized by  $\mathcal{H}_{f}=\otimes_{n}\mathcal{H}_{n}$, where $\mathcal{H}_{n}$ denotes the Hilbert space of mode $n$, spanned by the basis $\{|0_{n},\bar{0}_{n}\rangle,|1_{n},\bar{0}_{n}\rangle,|0_{n},\bar{1}_{n}\rangle,|1_{n},\bar{1}_{n}\rangle\}$, where the numbers of fermion and anti-fermion in mode $n$ are given by $j_{n}$ and $\bar{j}_{n}$, respectively. Thus the partition function can also be factorized by mode numbers as $Z=\prod_{n}Z_{n}$, where $Z_{n}=1+2e^{-\beta\omega_{n}}+e^{-2\beta\omega_{n}}$, and $\beta=\frac{1}{T_{R}}$. Thus the thermal state of the field can be written as
    \begin{equation}\label{eq:thermalstate}
        \rho_{0,f}=\bigotimes_{n}\sum_{j_{n}=0}^{1}\sum_{\bar{j}_{n}=0}^{1}P_{\{j_{n},\bar{j}_{n}\}}|j_{n},\bar{j}_{n}\rangle\langle j_{n},\bar{j}_{n}|,
    \end{equation}
    which is also a diagonal matrix, and the possibility of occupation state $j_{n},\bar{j}_{n}$ is given by $P_{\{j_{n},\bar{j}_{n}\}}=\frac{1}{Z}e^{-\beta\sum_{n}(j_{n}+\bar{j}_{n})\omega_{n}}$. It is easy to see that the trace of $\rho_{0,f}$ equals $1$. Taking $T_{R}\rightarrow 0$ only $P_{\{0,\bar{0}\}}$ survives thus making (\ref{eq:thermalstate}) the vacuum state. Note that in our setup the particles are boiled out of vacuum, thus the temperature of the field must be extremely high if we want to observe the existence of the particles with a reasonable probability when the mass $m$ of the particles is large.
    
    The anticommutation relation of the fermion operators $\{c_{n},c_{m}^{\dagger}\}=\delta_{nm},c=b,d$ makes it impossible to write $c_{n}^{\dagger}|0\rangle=|n\rangle,c_{n}^{\dagger}|n\rangle=0$ in a unified manner, thus the formula of heat transfer and entropy grow will be too complicated. However, things is not as hopeless as it seems. By observing (\ref{eq:WV}), we find that with under certain conditions, we can obtain an elegant formula approximately. When the Unruh-DeWitt detector stays still, and one of the field mode $B$ is in resonance with the detector, i.e. $\omega_{B}=\Omega$, also the interacting time $T$ is long enough, working out the integrals in (\ref{eq:WV}), we can see that $V_{B}$ will be the dominant part by growing proportional with the interaction time $T$ while the contribution of other terms have upper bounds.

    The initial state of the Unruh-DeWitt detector are chosen as $\rho_{0,D}=[(1-p)|-\rangle\langle-|+p|+\rangle\langle+|]$, whose direct product with (\ref{eq:thermalstate}) gives the state of the whole system. Use the same arguments in last section, we see $\rho^{(1)}(t)$ does not contribute to the heat transfer and entropy grow in this case either. To the second order of $\lambda$ and concentrate solely on the resonance mode $B$, we have
    \begin{equation}
        \begin{split}
            &U^{(1)}(T)(|1_{B},\bar{1}_{B}\rangle\langle 1_{B},\bar{1}_{B}|\otimes|-\rangle\langle -|)U^{(1)\dagger}(T)\rightarrow\lambda^{2}|V_{B}|^{2}|1_{B},\bar{0}_{B}\rangle\langle 1_{B},\bar{0}_{B}|\otimes|+\rangle\langle+|,\\
            &U^{(1)}(T)(|0_{B},\bar{0}_{B}\rangle\langle 0_{B},\bar{0}_{B}|\otimes|+\rangle\langle +|)U^{(1)\dagger}(T)\rightarrow\lambda^{2}|V_{B}|^{2}|0_{B},\bar{1}_{B}\rangle\langle 0_{B},\bar{1}_{B}|\otimes|-\rangle\langle -|.\\
            &U^{(1)}(T)(|1_{B},\bar{0}_{B}\rangle\langle 1_{B},\bar{0}_{B}|\otimes|+\rangle\langle +|)U^{(1)\dagger}(T)\rightarrow\lambda^{2}|V_{B}|^{2}|1_{B},\bar{1}_{B}\rangle\langle 1_{B},\bar{1}_{B}|\otimes|-\rangle\langle-|,\\
            &U^{(1)}(T)(|0_{B},\bar{1}_{B}\rangle\langle 0_{B},\bar{1}_{B}|\otimes|-\rangle\langle -|)U^{(1)\dagger}(T)\rightarrow\lambda^{2}|V_{B}|^{2}|0_{B},\bar{0}_{B}\rangle\langle 0_{B},\bar{0}_{B}|\otimes|+\rangle\langle+|,
        \end{split}
    \end{equation}
    and
    \begin{equation}
        \begin{split}
            &U^{(2)}(T)(|1_{B},\bar{0}_{B}\rangle\langle 1_{B},\bar{0}_{B}|\otimes|-\rangle\langle -|)=(|1_{B},\bar{0}_{B}\rangle\langle 1_{B},\bar{0}_{B}|\otimes|-\rangle\langle -|)U^{(2)\dagger}(T)\\
            &=-\frac{\lambda^{2}}{2}|V_{B}|^{2}|1_{B},\bar{0}_{B}\rangle\langle 1_{B},\bar{0}_{B}|\otimes|-\rangle\langle -|,\\
            &U^{(2)}(T)(|0_{B},\bar{0}_{B}\rangle\langle 0_{B},\bar{0}_{B}|\otimes|+\rangle\langle +|)=(|0_{B},\bar{0}_{B}\rangle\langle 0_{B},\bar{0}_{B}|\otimes|+\rangle\langle +|)U^{(2)\dagger}(T)\\
            &=-\frac{\lambda^{2}}{2}|V_{B}|^{2}|0_{B},\bar{0}_{B}\rangle\langle 0_{B},\bar{0}_{B}|\otimes|+\rangle\langle +|,\\
            &U^{(2)}(T)(|1_{B},\bar{0}_{B}\rangle\langle 1_{B},\bar{0}_{B}|\otimes|+\rangle\langle +|)=(|1_{B},\bar{0}_{B}\rangle\langle 1_{B},\bar{0}_{B}|\otimes|+\rangle\langle +|)U^{(2)\dagger}(T)\\
            &=-\frac{\lambda^{2}}{2}|V_{B}|^{2}|1_{B},\bar{0}_{B}\rangle\langle 1_{B},\bar{0}_{B}|\otimes|+\rangle\langle +|,\\
            &U^{(2)}(T)(|0_{B},\bar{1}_{B}\rangle\langle 0_{B},\bar{1}_{B}|\otimes|+\rangle\langle +|)=(|0_{B},\bar{1}_{B}\rangle\langle 0_{B},\bar{1}_{B}|\otimes|+\rangle\langle +|)U^{(2)\dagger}(T)\\
            &=-\frac{\lambda^{2}}{2}|V_{B}|^{2}|0_{B},\bar{1}_{B}\rangle\langle 0_{B},\bar{1}_{B}|\otimes|+\rangle\langle +|.\\
        \end{split}
    \end{equation}
    Tracing out the degrees of freedom of Dirac field and with the same notation in (\ref{eq:rhod}), we get
    \begin{equation}
            \delta p=\lambda^{2}|V_{B}|^{2}\bigg(1-p)P_{2}+(1-2p)P_{1}-pP_{0}\bigg),
    \end{equation}
    where
    \begin{equation}\label{eq:pn}
        \begin{split}
            &P_{0}=\sum_{\{j_{n'}\},\{\bar{j}_{n'}\};n'\neq B}P_{0_{B},\bar{0}_{B}},\\
            &P_{1}=\sum_{\{j_{n'}\},\{\bar{j}_{n'}\};n'\neq B}P_{1_{B},\bar{0}_{B}}=\sum_{\{j_{n'}\},\{\bar{j}_{n'}\};n'\neq B}P_{0_{B},\bar{1}_{B}},\\
            &P_{2}=\sum_{\{j_{n'}\},\{\bar{j}_{n'}\};n'\neq B}P_{1_{B},\bar{1}_{B}}.
        \end{split}
    \end{equation}
    Then we obtain the heat transfer and entropy grow as
    \begin{equation}\label{eq:thermalsq}
        \begin{split}
            &\Delta S=\lambda^{2}\ln{\frac{1-p}{p}}|V_{B}|^{2}\bigg(pP_{0}+(2p-1)P_{1}-(1-p)P_{2}\bigg),\\
            &\Delta Q=\lambda^{2}\omega_{B}|V_{B}|^{2}\bigg(pP_{0}+(2p-1)P_{1}-(1-p)P_{2}\bigg).
        \end{split}
    \end{equation}
    In this case, we can see the distinguishability of fermion and anti-fermion comes into play by shaping the structure of $\mathcal{H}_{n}$. Taking $T_{R}\rightarrow 0$, the results reduce to (\ref{eq:deltaqs}) under resonance approximation. Substitute (\ref{eq:pn}) into (\ref{eq:thermalsq}) it is easy to check that the Landauer's principle $\Delta Q\geq T_{R}\Delta S$ is satisfied. The initial state of the detector can also be assigned with a temperature $T_{D}$ with $p=\frac{e^{\frac{\omega_{B}}{T_{D}}}}{e^{\frac{\omega_{B}}{T_{D}}}+1}$.
    From (\ref{eq:thermalsq}) we obtain that the Dirac field will absorb heat when $T_{D}>T_{R}$ and emit heat when  $T_{R}>T_{D}$, which is in agreement with classical picture.

\section{Conclusion}\label{sec:5}

    In this work we studied the Landauer's principle of an Unruh-DeWitt detector coupled to massive and massless fermionic field in $1+1$ dimensional bag cavity. We have shown that when the Dirac field is prepared initially in vacuum state, the heat transferred to the Dirac field is always non-negative thus preserves the Landauer's principle. When the initial state of the Dirac field is a thermal state, we obtained the heat transfer and entropy grow approximately under some additional conditions and also verified the validity of the Landauer's principle. We find that compared to the case of real scalar field \cite{wyj1}, the results of vacuum initial state differs only due to the helicity of the Dirac field while for thermal initial state, the distinguishability of fermion and anti-fermion also plays its part. Our work holds the potential of being extended to the QCD systems and neucleaon-neutrino systems \cite{QCD,n1,n2}, which may be experimental feasible.
    
\section*{Acknowledgements}

\end{document}